\documentclass[9pt,twocolumn,aps,pre]{revtex4}
\usepackage{subfigure}
\usepackage{graphicx}
\usepackage[utf8]{inputenc}

\usepackage{float}
\usepackage{amssymb}

\begin{document}

\sloppy
\newcommand{\rs}{\begin{equation}}
\newcommand{\re}{\end{equation}}
\newcommand{\eas}{\begin{eqnarray}}
\newcommand{\eae}{\end{eqnarray}}

\renewcommand{\subfigtopskip}{0.1cm}
\renewcommand{\subfigcapskip}{0cm}
\renewcommand{\subfigbottomskip}{0cm}
\renewcommand{\subfigcapmargin}{0cm}

\title{Thermal diffusion of solitons on anharmonic chains with long-range
coupling}
\author{C. Brunhuber}    
\email{Christian.Brunhuber@uni-bayreuth.de}
\author{F.G. Mertens}
\affiliation{Physikalisches Institut, Universit\"at Bayreuth, D-95440 Bayreuth, Germany}
\author{Y. Gaididei}
\affiliation{Bogolyubov Institute for Theoretical Physics, 03143 Kiev, Ukraine}

\begin{abstract}
We extend our studies of thermal diffusion of non-topological solitons to
anharmonic FPU-type chains with additional long-range couplings. 
The observed superdiffusive behavior in the case of nearest neighbor
interaction (NNI) turns out to be the dominating mechanism for the soliton diffusion on chains with  long-range interactions (LRI). 
Using a collective variable technique in the framework of a variational
analysis for the continuum approximation of the chain, we derive a set of stochastic integro-differential equations for the collective variables (CV) soliton position and the
inverse soliton width.
This set can be reduced to a statistically equivalent set of Langevin-type
equations for the CV, which shares the same Fokker-Planck equation.
The solution of the Langevin set and the Langevin dynamics simulations of the discrete system agree well and 
demonstrate that the variance of the soliton increases stronger than linearly with time
(superdiffusion). This result for the soliton diffusion on anharmonic chains
with long-range interactions reinforces the conjecture that superdiffusion is a generic
feature of
non-topological solitons.
\end{abstract}

\maketitle
\section{Introduction}
Solitons appear in various different fields of physics such as solid state
physics, quantum theory, nonlinear optics, fluid dynamics, biophysics or even
cosmology \cite{scott,encyclopedia}. The numerical
experiment done by Fermi, Pasta and Ulam in 1955, investigating the thermalization of a
one-dimensional chain with anharmonic interaction potential, was certainly one
of the key events in nonlinear science \cite{fermi}.
Hereafter, FPU chains, despite of their simplicity, served as frequently used
models for analyzing the properties of microscopic structures, especially in computer simulations. 
More than fifty years after the above mentioned experiment by Fermi, Pasta and
Ulam, complete understanding of this putatively simple system, which still
reveals new astonishing phenomena \cite{campbell05}, is yet to be reached. The thermalization of
the FPU system, the heat conduction in FPU systems or the properties of lattice
solitons and discrete breathers are still tasking many physicists (see for
example CHAOS
Volume 15: focus issue: THE "FERMI-PASTA-ULAM" PROBLEM-THE FIRST 50 YEARS).
\\
Physicists usually use simplified models to describe and explain the most
important features of more complex systems. An example is the "invasion" of physicists into
biology, where they aimed to explain some of the most interesting processes in
living cells such as the DNA dynamics (DNA melting, DNA transcription) or the
functionality of proteins as molecular motors, and the storage and transport of
energy released by the hydrolysis of adenosine-triphosphate (ATP) \cite{peyrard04,yakushevic01,dauxois1,scott92,mingaleev99,choi04}. Oversimplified
models, which were formerly used to predict freely moving solitons along DNA or protein
$\alpha$-helices may be too rough to explain the dynamics of biomolecules. But,
for example, 
taking the role of discreteness, thermal perturbations
and the underlying geometry of the system into account, it is possiblee to obtain good agreements with the
results of experiments, which today work at the scale of single molecules. 
\\
The Davydov model \cite{davydov,scott92} for energy transport and storage in
proteins has attracted much attention in the past. Davydov proposed a self-trapping of vibrational
energy of the C=O vibration, similar to the concept of polarons in solid-state
physics. 
Recently, a direct observation of
self-trapped states in pump-probe experiments of a model protein (ACN), 
showed that the self-trapping could describe some important features of proteins
(although the lifetime of
the excitation in isolated molecules is shorter than expected by 
Davydov) \cite{edler02}.
In a different approach by Yomosa \cite{yomosa85}, a molecular mechanism of muscle contraction
was introduced, which
assumes that the H-bonding interaction of the peptide groups (which stabilize the
helix structure) can be modeled by a FPU-type lattice. The energy transport
along the helix was described in the continuum approximation by use of the
stability of the Korteweg-de Vries (KdV) or the Toda soliton \cite{perez86,perez87,hochstrasser89}. Although the continuum approximation and the
disregard of the helical structure seems to be bold to model real processes,
the concept of lattice solitons in protein backbones was further refined. It was
numerically demonstrated that protein backbones with a more realistic structure,      
support stable non-topological solitons \cite{christiansen97}. Recently, the concept of solitons along
hydrogen bonds in $\alpha$ helices was applied to realistic models for proteins
in order to explain Raman measurements, which identify excitations in proteins, 
which can not stem from vibrational modes in the molecule and are too long-lived
to result from large-scale vibrations of the protein \cite{dOvidio03,dOvidio05}. The three-dimensional model of the
 $\alpha$ helix consists of three chains with hydrogen bonds (secondary
 structure), which are connected with peptide bonds (primary
 structure), modeled by Lennard-Jones or Toda potentials. It was demonstrated that a
 perturbation in one of the three H-bond lines can create a triple-soliton in the
 three hydrogen bonds, which travels (phase-locked by the peptide bonds) along the
 chain and would present a long-lived energy source for the spectroscopically observed phonons at 
 energies of about $ \sim 100~ cm^{-1}$.    
\\ 
Often, the dynamics of more realistic models can only be examined or dealt with
in computer simulations. Yet, sometimes simplified models, such as the
FPU-chain, are able to provide proper understanding of these models.
Therefore, it is important to clarify the effect of extensions on systems such
as FPU-chain and if possible, apply analytical calculations. 
In the case of biomolecules, especially damping and thermal fluctuations appear as natural perturbations. 
The influence of the thermal fluctuations on solitons on FPU chains
(with cubic or quartic anharmonicity) has recently been investigated \cite{arevalo03,brunhuber04,mertens05,arevalo02}.
It was demonstrated that both types of non-topological solitons, pulse solitons and
envelope solitons, show a
position variance growing stronger than linearly in time (superdiffusion),
whereas the mechanisms and the time-dependencies differ.
Envelope solitons are generally more sensitive to damping than pulse solitons.
This sensitivity leads to
an exponentially growing position variance, which ends when the envelope soliton is destroyed by damping and noise. The pulse solitons
have a much longer lifetime and the time-dependence of the position variance
proved to have a linear (normal) and a quadratic (superdiffusive) part \cite{arevalo03}.
The superdiffusion appears to be typical for non-topological solitons
because, whereas
it was also observed for pulse solitons in classical Heisenberg systems
\cite{kamppeter99,meister01}, 
it is absent in the case of topological solitons in the sine-Gordon system \cite{quintero00}.
\\
In this article, the aim is to extend the results to chains with long-range
interactions (LRI). LRI are necessary extensions which help make the models more
realistic. In biomolecules, charged groups or dipole moments (for example the dipole moment of the base pairs in DNA) should have an effect on solitonic or discrete
excitations \cite{mingaleev99,archilla02,mingaleev98}. In the case of the FPU
system, it was demonstrated that chains with Kac-Baker or power-law long-range interactions support
pulse solitons as well \cite{mingaleev98,mingaleev00}. In the case of Kac-Baker
LRI, two different velocity branches for soliton solutions, which are separated
by a gap, appear. Low-velocity solutions with a soliton width larger than the interaction
radius are pulse-shaped, but develop a cusp when the soliton velocity is close to the critical velocity at the gap \cite{neuper94,gaididei97}. High-velocity solitons with a width in the range of the
interaction radius have a multi-component structure, where the center of the
soliton is mainly determined by the NNI, whereas the tails are determined by
the LRI \cite{mingaleev00,christiansen01}.         
\\ 
According to our research, this is the first known attempt to study the effect
of thermal fluctuations on solitons, of wich the shapes are governed by long-range interactions. Our goal is to
investigate the consequences of the long-range interactions on the stochastic
behavior of the soliton, and specifically on the time-dependence of its
position variance.

\section{The Model}
Our model is a one-dimensional chain of equally spaced particles of mass m
($m=1$) with interatomic spacing $a$ ($a=1$). The displacement of the particle $n$ from its
equilibrium position is denoted as $u_n$.
The potential consists of a part $U_{NN}$, with only NNI similar to the original
potential chosen by Fermi, Pasta and Ulam, and a long-range part $U_{LR}$ with harmonic coupling $J_{nm}$ between the
particles $n$ and $m$:  
\eas
L=T-U_{NN}-U_{LR}
\eae
with
\eas
T=\frac{1}{2} \sum_n \bigg( \frac{d u_n}{d t}\bigg)^2 \;,\;U_{NN}=\sum_n V
(u_{n+1}-u_n) \nonumber \\
U_{LR}=\frac{1}{2} \sum_{n,m} J_{n,m} (u_n-u_m)^2~~.
\eae
The potential $V(r)$ between nearest neighbors represents the expansion of the Toda potential to third order
\eas
V(r)=\frac{1}{2} r^2-\frac{1}{3} r^3~~.
\label{NNI}
\eae
The long-range coupling between sites $n$ and $m$ is of the Kac-Baker form \cite{baker61,kac72}, $\alpha^{-1}$ is the interaction radius of the LRI
\eas
 J_{n-m} = \frac{1}{2} J
(e^{\alpha}-1) e^{-\alpha \mid n-m\mid}~,~\sum_{m}^NJ_m=J~.
\label{Jsum}
\eae
For modeling the environment, we choose hydrodynamical damping instead of Stokes
damping (because it was demonstrated in \cite{arevalo02} that
the long-wave components of the Fourier spectrum are strongly damped for Stokes
damping which leads to deformations of the soliton) and a Gaussian white noise term which
fulfills the fluctuation-dissipation-theorem ($D^{hy}=2 \nu_{hy} T$ with $k_B$ set to unity). The equations of motion in relative displacement coordinates $w_n=u_{n+1}-u_n$ then take the form
\eas
\ddot{w}_n= & & V^{\prime}(w_{n+1}) -2 V^{\prime}(w_{n})+
V^{\prime}(w_{n-1}) \nonumber\\
&-&\frac{J}{2} (e^{\alpha}-1) \sum_{m} (w_n -w_{n+m}) e^{-
\alpha \mid m \mid}
\nonumber\\
&+&\nu_{hy} (\dot{w}_{n+1}-2
\dot{w}_n+\dot{w}_{n-1}) \nonumber \\
&+& \sqrt{D^{hy}} (\xi_{n+1}(t)-\xi_n(t))~.
\label{discreteeq}
\eae
In order to apply similar analytic tools as used in \cite{arevalo03,brunhuber04}, we follow Ref.\cite{gaididei97} and derive in the quasi-continuum approximation (QCA) 
a partial differential equation (PDE), which is of the Boussinesq (Bq) -type  
\eas
\partial_t^2 w=(c^2-1) \frac{\kappa^2 \partial_x^2}{\kappa^2-\partial_x^2}
w+\partial_x^2 (w-w^2)+\lambda \partial_x^2 \partial_t^2 w\nonumber \\+\nu_{hy}
\partial_x^2 \partial_t w + \sqrt{D^{hy}} \partial_x^2 \xi(x,t)
\label{conteq}
\eae
\eas
\lambda=\frac{1}{12}~,~\kappa=2 sinh \bigg( \frac{\alpha}{2} \bigg)
\\
c=\bigg[1+ \frac{1}{2} J \frac{e^{-\alpha}+1}{(e^{-\alpha}-1)^2}
\bigg]^{\frac{1}{2}}
\eae
with an additional damping, noise and long-range term (a pseudo differential operator) which is defined as
\eas
(\alpha^2-\partial_x^2)^{-1}f(x)=\frac{1}{2 \alpha} \int_{-\infty}^{\infty} dx^{\prime} e^{-\alpha \mid x-x^{\prime} \mid} f(x^{\prime})
~~.\eae

\begin{figure}[htbp]
\caption{(Color online) Soliton solutions of eq. (\ref{conteq}) with the same normalized velocity $c_o=v/c=1.01$ for a chain with nearest neighbor interactions ($J=0$)
and for a chain with a relatively weak long-range coupling $J=0.1$ for different values of the interaction radius $\alpha^{-1}$.}
\hspace*{-0.5cm}\subfigure{\includegraphics[width=7cm,angle=270]{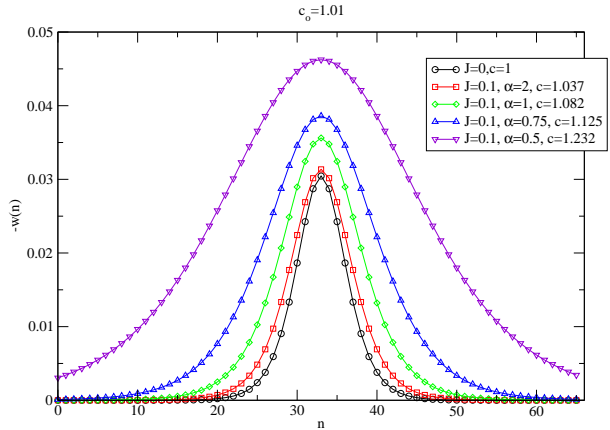}}
\label{nnwwlrwwvergl}
\end{figure}

In \cite{gaididei97,mingaleev00}, the solution of equation (\ref{conteq}) (for $\nu_{hy}=0$ and $D^{hy}=0$) was given implicitly. 
It was demonstrated that the chain with Kac-Baker LRI supports soliton solutions in the velocity interval $c<v<v_c$,
where $c$ is the velocity of sound and $v_c \simeq \sqrt{(4 c^2-1)/3}$ is the
critical velocity, where the solitons develop a cusp like peakons, which are
known from the theory of shallow water waves \cite{whitham} and the integrable Camassa-Holm equation \cite{camassa93}. A stability analysis in \cite{gaididei97} proved
that the soliton becomes unstable before it adopts peakon shape. It was further
demonstrated that it is possible to construct long-range potentials in nonlinear lattices
which support (discrete) peakon solutions \cite{comech05}. 
\\
Here, the aim is not to directly solve equation (\ref{conteq}), but rather to
benefit from the fact that as long as the soliton velocity is smaller than $v_c$, the soliton can be
relatively well approximated by a pulse-shaped collective variable (CV) ansatz. 
As demonstrated in Fig. \ref{nnwwlrwwvergl}, an increasing long-range radius
$\alpha^{-1}$ yields larger and broader
excitations even when the long-range coupling $J$ is relatively small. For
$\alpha=0.5$ and $J=0.1$, the soliton with the normalized velocity $c_o=v/c=1.01$ is much broader than the
soliton solution without long-range forces ($J=0$) with the same $c_o$.
Therefore, the continuum approximation seems to be a very appropriate
analytical tool to investigate long-range effects.     

\begin{figure}[htbp]
\caption{(Color online) Soliton solution for different values of $c_o$ for $J=0.1$ and $\alpha=0.3$. For $c_o=v_c/c=1.092$, the soliton solution takes the form of a peakon.}
\hspace*{-0.5cm}\subfigure{\includegraphics[width=7cm,angle=270]{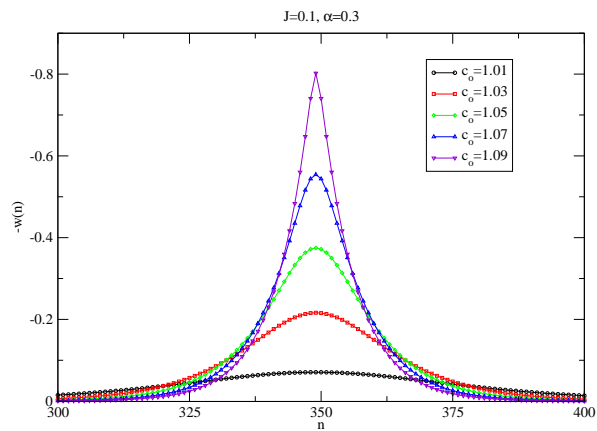}}
\label{vvergl}
\end{figure}
The case of nearest neighbor interactions is covered by our theory in the limit
$J\rightarrow 0$ (for $\alpha \neq 0$). Here, the long-range term in equation (\ref{conteq})
disappears when $(c^2-1)\rightarrow 0$ and we obtain an improved Boussinesq
equation (IBq) with $c=1$ \cite{arevalo03}. The limit $\alpha \rightarrow \infty$ leads to a IBq
with an additional nearest neighbor coupling $J/2$ and a sound velocity
$c=\sqrt{1+J/2}$.

\section{Collective Variables}
A collective variable theory depends crucially on the trial function one chooses to derive ordinary differential equations (ODEs) for the 
soliton parameters, such as width and position. As seen in Fig. \ref{vvergl}, the shape of the soliton is rather pulse-like up to values $c_o\le1.07$. A  CV ansatz of this form 
has already achieved qualitatively good results with simulations for a large range of possible soliton velocities \cite{neuper94}. In absolute displacement coordinates $u(x,t)$, the
pulse-shaped soliton solution (in relative displacement coordinates $w(x,t)=\partial_x u(x,t)$) takes the form of a kink, where the amplitude 
\eas
2 A_o=\int_{-\infty}^{\infty} w(x,t) dx
\eae  
is a conserved quantity (such as for the Bq equation). Therefore, we choose a trial function of the form
\eas
u(x,t)=A_o tanh \big( \gamma(t) (x-X(t)) \big)
\label{ansatz}
\\
w(x,t)=\partial_x u(x,t)=A_o \gamma(t) sech^2 \big( \gamma(t) (x-X(t) \big)
\eae
with the time-dependent collective coordinates $\gamma(t)$ (inverse width) and $X(t)$ (position) and the kink amplitude $2 A_o$ of the unperturbed initial soliton. 
This CV ansatz yields the correct results (the Bq soliton solution) in the case of an unperturbed system with only NNI (\ref{NNI})
\eas
\gamma=\sqrt{3} \frac{\sqrt{v^2-1}}{v}~,~A_o=\frac{\sqrt{3}}{2} \sqrt{v^2-1} v\nonumber \\
X(t)=v t+X_o~~.
\eae
\\
In order to derive ODEs for the collective variables $\gamma(t)$ and $X(t)$, we apply a variational technique similar to the method of Whitham \cite{whitham}.
It is well known that a Lagrange density of Boussinesq-type equations only exists for the absolute coordinate field $u(x,t)$ \cite{jeffrey}. In order to include the damping term, 
the generalized Hamilton principle of Ostrovsky et al.
\cite{ostrovsky71,jeffrey}, who extended the method of Whitham to
nonconservative systems, is here applied.
The equation (\ref{conteq}) in absolute displacement coordinates 
\eas
\partial_t^2 u&-&\lambda \partial_x^2 \partial_t^2 u-(c^2-1) \frac{\kappa^2 \partial_x^2}{\kappa^2-\partial_x^2}
u-\partial_x^2 u+2 (\partial_x u) (\partial_x^2 u) \nonumber \\&-&\sqrt{ D^{hy} } \partial_x
\xi(x,t)  = \nu_{hy} \partial_x^2 \partial_t u
\;\;.\eae
can be written as 
\eas
-\frac{\partial \mathcal{L}}{\partial u}+
\frac{\partial}{\partial t}\frac{\partial \mathcal{L}}{\partial u_t}
+\frac{\partial}{\partial x}\frac{\partial \mathcal{L}}{\partial u_x}
-\frac{\partial^2}{\partial x \partial_t} \frac{\partial
\mathcal{L}}{\partial u_{x t}}=\Phi
\label{Ldensityrule}
\eae
with
\eas
\mathcal{L}=\frac{u_t^2}{2}-\frac{u_x^2}{2}+\frac{u_x^3}{3}+\frac{\lambda}{2}
u_{x t}^2\nonumber +\sqrt{D^{hy}} u_x \xi(x,t)\nonumber\\+\frac{(c^2-1) \kappa}{4} \int_{-\infty}^{\infty} e^{- \kappa \mid
x-s \mid}  u(x,t)\partial_s^2 u(s,t) ds
\label{Ldensity}
\\
\Phi=\nu_{hy} \partial_x^2 \partial_t u~.
\eae
Therefore, the generalized Hamiltonian principle for wave motion in nonlinear and dissipative media 
in the framework of our CV theory, is used here. Accordingly, the equations of motions can be derived from the averaged Lagrangian density and damping term by varying with
respect to the CV ($X_1=\gamma(t), X_2=X(t)$) 
\eas
\frac{\delta <\mathcal{L}>}{\delta X_i}=\frac{\partial <\mathcal{L}>}{\partial X_i}-\frac{\partial
}{\partial t} \frac{\partial <\mathcal{L}>}{\partial \dot{X}_i}=- \bigg< \Phi
u_{i} \bigg>~.
\label{Leq}
\eae
The brackets signify a spatial integration over $x$.
In \cite{brunhuber04}, where the soliton equation was the nonlinear
Schr\"odinger equation, it was possible to include the damping in the Lagrangian
density and, thus,  no damping term
$\Phi$ was needed. Further, the treatment of the damping for the
KdV \cite{arevalo03}, the treatment of the damping proved unproblematic, because
a projection method was used for the CV \cite{mertens97} (this method is,
however, rather extensive for the
non-local Bq equation (\ref{conteq})).

\section{Langevin equations}
Substituting the ansatz (\ref{ansatz}) in the Lagrangian density
(\ref{Ldensity}) and applying the variation (\ref{Leq}), results in
the following set of stochastic
integro-differential equations, where small terms , $\sim \dot{\gamma}^2$, $\sim
\ddot{X}$ and $\sim \ddot{\gamma}$ were neglected to achieve analytically
manageable equations ($\theta=\kappa/2 \gamma$, $\bar{x}=x-X(t)$)
\eas
&&\frac{4}{3}\dot{\gamma} \dot{X}-\frac{2 \sqrt{D^{hy}} \gamma^2}{A_o}
\int^{\infty}_{-\infty} Sech^2[\gamma \bar{x}] Tanh[\gamma \bar{x}] \xi(\bar{x},t)  d\bar{x}=\nonumber \\&&-\frac{16}{15} \nu_{hy} \gamma^3 \dot{X}
\label{CVnoisy1}
\eae
\eas
&-&2(\dot{X}^2-1)+\frac{16}{9} A_o \gamma-4 (c^2-1) 
\bigg[\theta+\theta^2+\theta^4
\Psi^{\prime \prime}(\theta) \bigg]
\nonumber\\
&+&\frac{2 \sqrt{D^{hy}} \gamma }{A_o}
\int_{-\infty}^{\infty} d \bar{x} Sech^2[\gamma \bar{x}] Tanh[\gamma \bar{x}]  
\bar{x} \xi(\bar{x},t)
\nonumber \\
&-&\frac{2 \sqrt{D^{hy}}}{A_o}
\int_{-\infty}^{\infty} d \bar{x} Sech^2[\gamma \bar{x}]  
\xi(x,t)=0~~.
\label{CVnoisy2}
\eae
The values for higher time derivatives of $\gamma(t)$ and $\dot{X}$ can be estimated from the simulation results. A classification
of the contributions in orders of a small parameter $\epsilon \sim \gamma_o$ makes
obvious that these terms contribute only negligibly. It was additionally
checked, that the numerical solution of system (\ref{CVnoisy2}) \cite{wolfram} without noise yields practically the same results
as the complete set without neglecting the small terms. At this point, it should
be mentioned that the neglect of $\ddot{X}$ is justified by very
different arguments compared to those for systems of point particles in the
overdamped limit, for instance in ratchet systems. In the following
calculations, it will be demonstrated that the dynamics of the
system depend on the reduced time scale $t_r=\gamma_o^2 \nu_{hy}t$. Therefore, time derivatives appear in an order $\mathcal{O}(\epsilon^4)$ smaller because we choose $\nu_{hy}$ in the order
$\mathcal{O}(\epsilon^2)$.
\\
The section in the square brackets of (\ref{CVnoisy2}) results from the long-range interactions, where $\Psi^{\prime \prime}$ is the second derivative of the digamma function.
In order to get a first-order equation in $X(t)$, we rewrite equation (\ref{CVnoisy2}) to  
\eas
\dot{X}^2=v_d^2 \bigg(1+\frac{1}{v_d^2}  \bigg( -\frac{2 \sqrt{D^{hy}}}{A_o}
\int_{-\infty}^{\infty} d x Sech^2[\gamma \bar{x}]  
\xi(x,t)+\nonumber \\
+\frac{2 \sqrt{D^{hy}} \gamma }{A_o}
\int_{-\infty}^{\infty} d x Sech^2[\gamma \bar{x}] Tanh[\gamma \bar{x}] \bar{x} 
\xi(x,t)  \bigg) \bigg)
\;\;.
\label{rewritten}
\eae
with
\eas
v_d^2=1+\frac{8}{9} A_o \gamma(t)-2(c^2-1) \bigg[
\theta+\theta^2+\theta^4
\Psi^{\prime \prime}(\theta) \bigg]~.
\label{vd}
\eae
For zero temperature, $v_d$ is the velocity of the damped soliton where the
damping effects enter in the form of the soliton width $\gamma(t)$. 
To investigate the
influence of the noise, the equations for the CV need to be written as
Langevin-type equations. Thus, we approximate $\dot{X}$ in (\ref{CVnoisy1}) and
$v_d$ in equation
(\ref{rewritten}) with the initial velocity of the soliton $v$, extract the
square root of (\ref{rewritten}) and approximate the r.h.s. by expanding the root to the first order in
the small parameter $\sqrt{D^{hy}}$. The result is a system of stochastic integro-differential equations for the CV:

\eas
\left( \begin{array}{c} \dot{\gamma} \\ \dot{X} 
\end{array} \right)  = 
\left( \begin{array}{c} A_1\\ A_2
\end{array} \right) + \int_{-\infty}^{\infty} dx \left( \begin{array}{cc}  
B_{11}&0 \\ B_{21}&B_{22}  \end{array} \right)
\left( \begin{array}{c} \xi \\ \xi \end{array} \right)
~,
\label{sideq}
\eae
with
\eas
A_1=-0.8 \nu_{hy} \gamma^3 ~~, ~~A_2=v_d
\\
B_{11}=\frac{3}{2} \frac{\sqrt{D^{hy}\gamma^2}}{A_o v} Sech^2 \big[\gamma
\bar{x} \big] Tanh \big[ \gamma
\bar{x}  \big]
\\
B_{21}=-\frac{\sqrt{D^{hy}}}{2 A_o v} Sech^2 \big[ \gamma
\bar{x} \big]
\\
B_{22}=\frac{\sqrt{D^{hy}} \gamma}{2 A_o v} Sech^2 \big[ \gamma
\bar{x} \big] Tanh \big[ \gamma
\bar{x}\big] \bar{x}~~.
\eae
As in \cite{arevalo03,brunhuber04}, we proceed to find a statistically equivalent Langevin system (yielding the same
Fokker-Planck equation as (\ref{sideq}) in the Stratonovich interpretation
\cite{konotop}), which is more convenient for further numerical and analytical studies. The
Langevin-system with two independent Gaussian white noise processes reads: 
\eas
\left( \begin{array}{c} \dot{\gamma} \\ \dot{X} 
\end{array} \right) =\left( \begin{array}{c} a_1\\ a_2
\end{array} \right)+\left( \begin{array}{cc}  
b_{11}&0 \\ 0&b_{22}  \end{array} \right)
\left( \begin{array}{c} \xi^1 \\ \xi^2 \end{array} \right)
\label{langevin}
\eae
with
\eas 
a_1=A_1+\frac{1}{4} \frac{D^{hy} }{A_o^2 v^2} \gamma^2 ~,~a_2=v_d
\nonumber \\
b_{11}=\sqrt{\frac{3}{5}} \frac{\sqrt{D^{hy}} }{A_o v} \gamma^{\frac{3}{2}}
\nonumber \\
b_{22}=\sqrt{(\frac{1}{6}+\frac{\pi^2}{180})}  \frac{\sqrt{D^{hy}} }{A_o v \gamma^{\frac{1}{2}}}~.
\eae  
\\
At this point, it is possible to compare the Langevin system here presented,
with the results only for NNI in reference \cite{arevalo03}. Despite the facts that the soliton diffusion
in \cite{arevalo03} was calculated
for low-velocity solitons with a CV ansatz representing a KdV soliton, and that
a different CV procedure was used, interesting
similarities can be observed. The stochastic equation for $\gamma(t)$ in
(\ref{langevin}) demonstrates, as in \cite{arevalo03}, a damping induced broadening of the soliton where the drift term is
proportional to the damping constant and to the third power of $\gamma$
($\dot{\gamma}\sim \nu_{hy} \gamma^3$), plus a small correction which is proportional to $D^{hy}$.
The effect of the damping, namely the soliton broadening mechanism, seems to be
the same for solitons of the KdV and the Bq type. However, and more importantly,
the long-range interactions
do not seem to change this behavior (yet, of course they change the initial condition $\gamma_o=\gamma(t=0)$ of the soliton). 
\\
In Ref. \cite{arevalo03}, it is argued that the observed superdiffusive behavior is induced by the stochasticity of $\gamma(t)$ and its effect on the soliton velocity. The
fluctuations of the soliton shape lead to fluctuations of the soliton velocity,
and therefore, to the superdiffusive terms in the position variance. For
envelope solitons, however, it was
shown that the superdiffusive mechanism differs. The reason is that the rapid
broadening of the envelope due to the damping, and not the changes due to
fluctuations, is the main contribution \cite{brunhuber04}. 
\\
If it is assumed that, as for envelope solitons \cite{brunhuber04}, the fluctuations of $\gamma(t)$ are negligible and the main contribution to the soliton diffusion results from the noise acting
directly on the soliton center $X(t)$, $b_{11}=0$ can be set and we can use the analytical
result for the soliton width, which depends on the reduced time $t_r=\gamma_o^2
\nu_{hy}t$ 
\eas
\gamma(t)=\frac{\gamma_o}{\sqrt{1.6 t_r + 1}}
\label{gamma}
\eae
to calculate the position variance
\eas
Var[X]_{b_{11}=0}=\int_{0}^t b_{22}^2(t^{\prime})
dt^{\prime}
\nonumber\\
=\bigg(\frac{1}{6}+\frac{\pi^2}{180} \bigg) \frac{D^{hy}}{A_o^2 v^2}
\int_{0}^t \gamma^{-1}(t^{\prime}) d t^{\prime}
\nonumber\\
=\frac{k_B T}{A_o^2 v^2} \frac{5 (\frac{1}{6}+\frac{\pi^2}{180})}{6}  \bigg(\frac{1}{\gamma(t)^3}- \frac{1}{\gamma_o^3} \bigg)
~~.\eae
Because higher order terms do not appear until $t_r \lesssim 1$, this result
yields an approximately linear time-dependence for typical simulation times. 
Therefore, $Var[X]_{b_{11}=0}$ will be denoted as the normal diffusion of the
soliton below. The result $Var[X]_{b_{22}=0}$ which we obtain by solving numerically the Langevin system (\ref{langevin})
with the constraint $b_{22}=0$ bears only the stochastic contribution of
$\gamma(t)$ and can be regarded as the anomalous or superdiffusive part of the soliton position. Although the equality $Var[X]=Var[X]_{b_{11}=0}+Var[X]_{b_{22}=0}$ does not
hold, these
two quantities can be used to illustrate the weight of normal and superdiffusive effects.

\section{Simulations}
\begin{figure}[htbp]
\caption{Soliton diffusion for different solitons ($c_o=1.03$, $c_o=1.05$ and $c_o=1.07$) on a chain with $J=0.1$, $\alpha=0.3$, $\nu_{hy}=0.01$ and $T=0.0001$.
The results $Var[X]_{b_{11}=0}$ and $Var[X]_{b_{22}=0}$ are obtained by solving the system (\ref{langevin}) for $b_{11}=0$ and $b_{22}=0$. The result for $Var[X]_{b_{22}=0}$ illustrates the dominance of the
superdiffusive mechanism.
}
\subfigure[]{\includegraphics[width=7cm,angle=270]{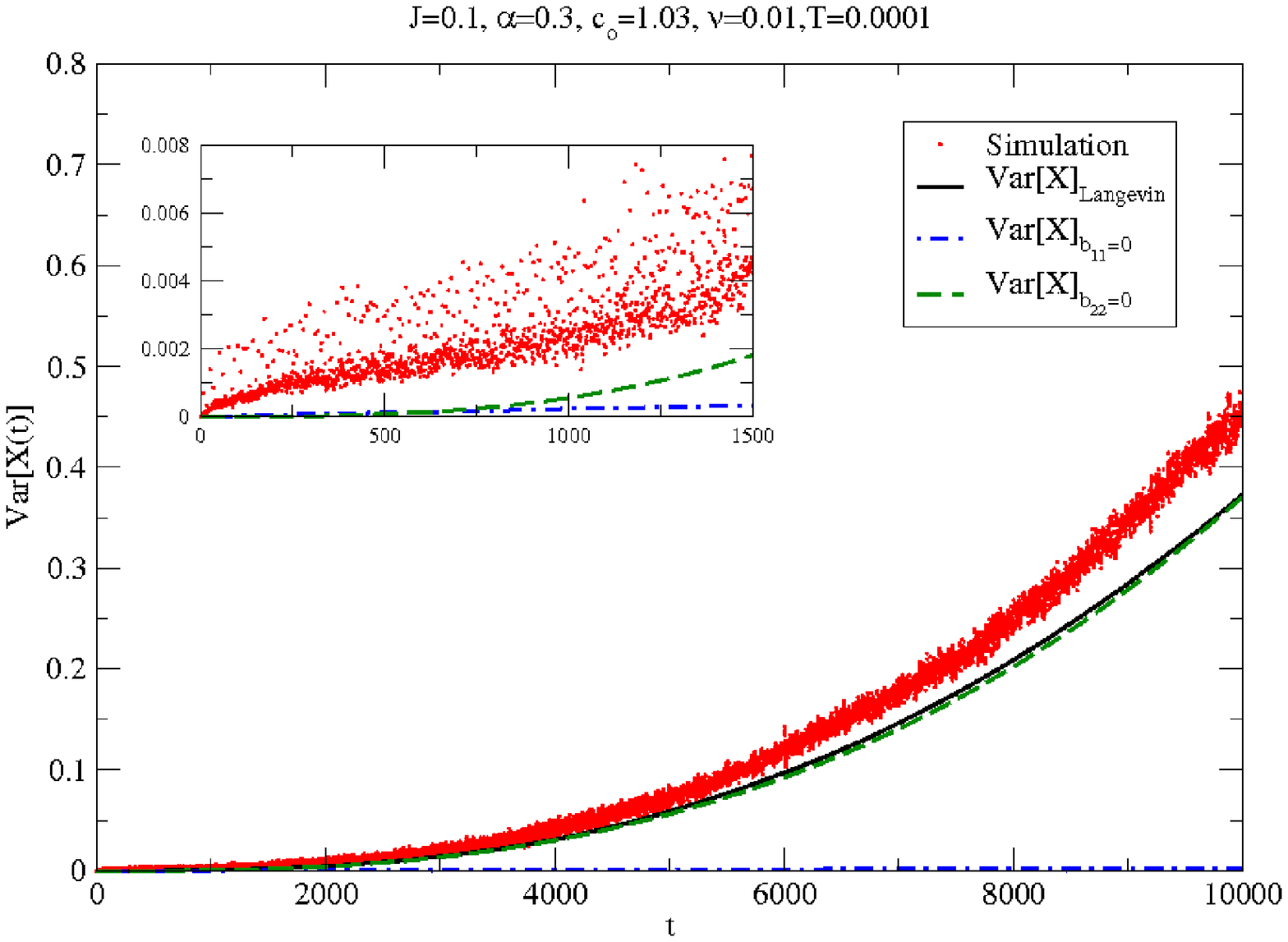}}
\subfigure[]{\includegraphics[width=7cm,angle=270]{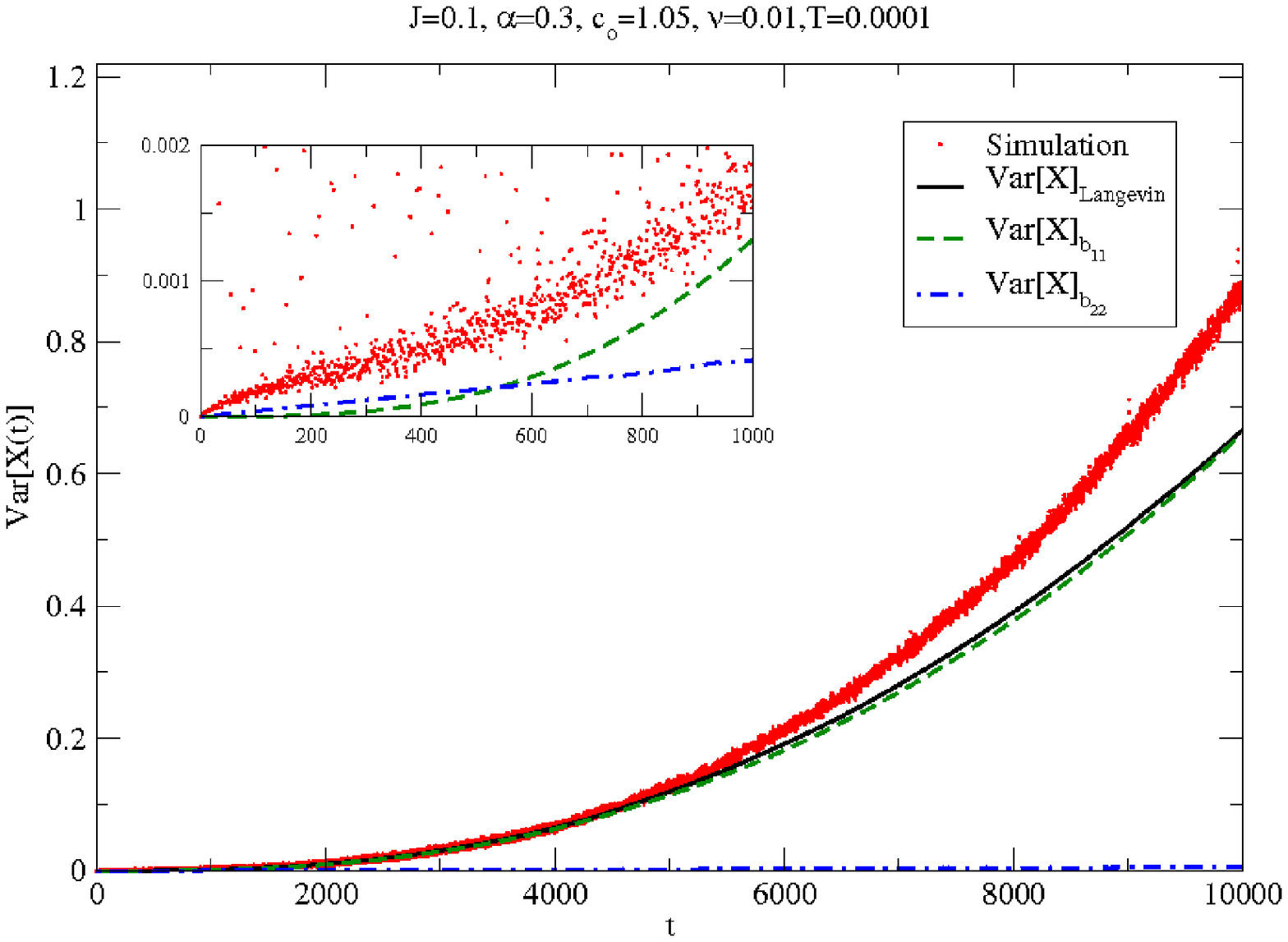}}
\subfigure[]{\includegraphics[width=7cm,angle=270]{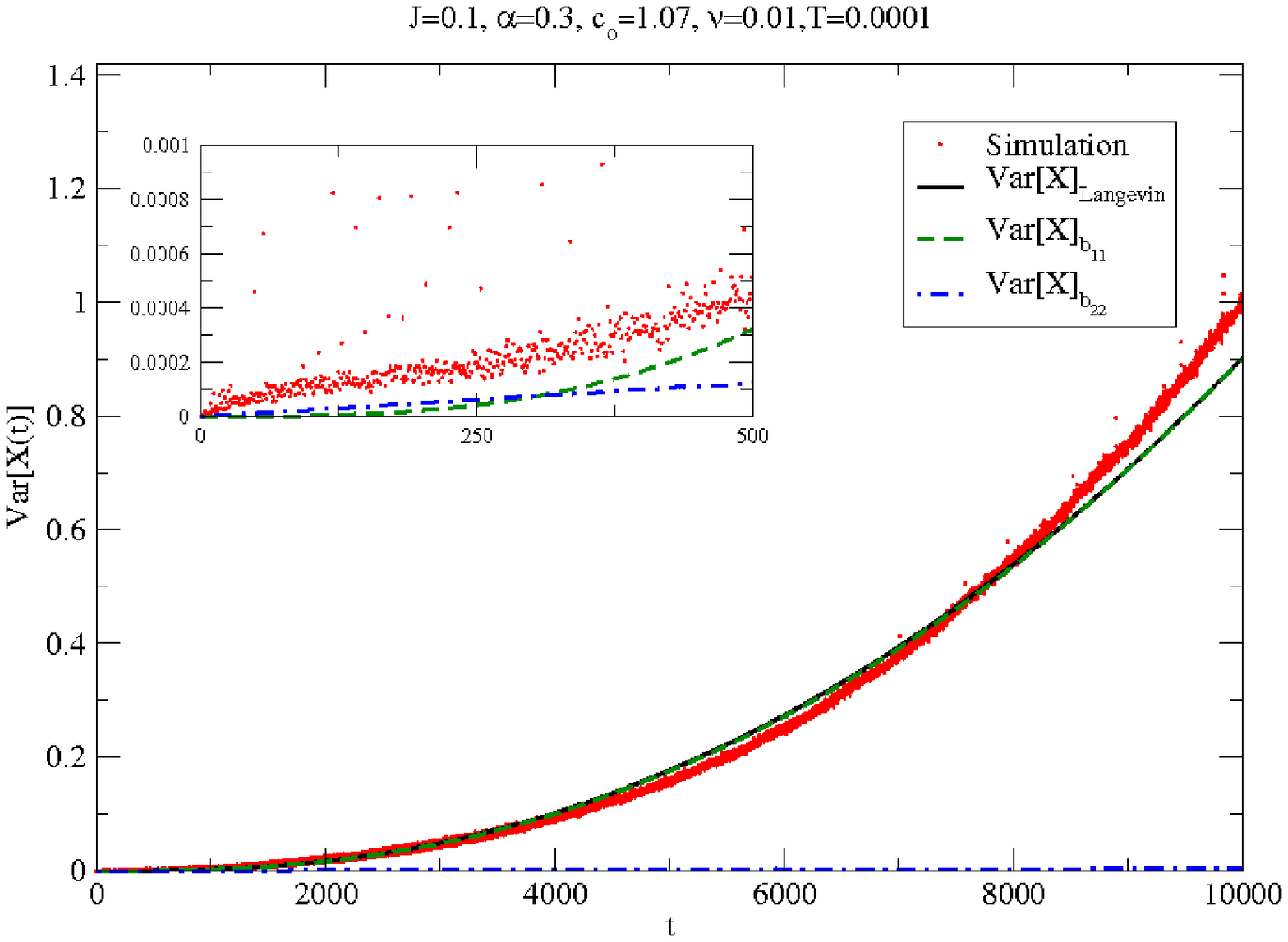}}
\label{soldiffal0.3}
\end{figure}

\begin{figure}[htbp]
\caption{Soliton diffusion for different solitons ($c_o=1.03$, $c_o=1.05$ and $c_o=1.07$) on a chain with $J=0.0468$, $\alpha=0.2$, $\nu_{hy}=0.01$ and $T=0.0001$. 
The results $Var[X]_{b_{11}=0}$ and $Var[X]_{b_{22}=0}$ are obtained by solving the system (\ref{langevin}) for $b_{11}=0$ and $b_{22}=0$. The result for $Var[X]_{b_{22}=0}$ illustrates the dominance of the
superdiffusive mechanism. The small off-set in (a) results from the inaccuracy of measuring the soliton position for very broad solitons.}
\subfigure[]{\includegraphics[width=7cm,angle=270]{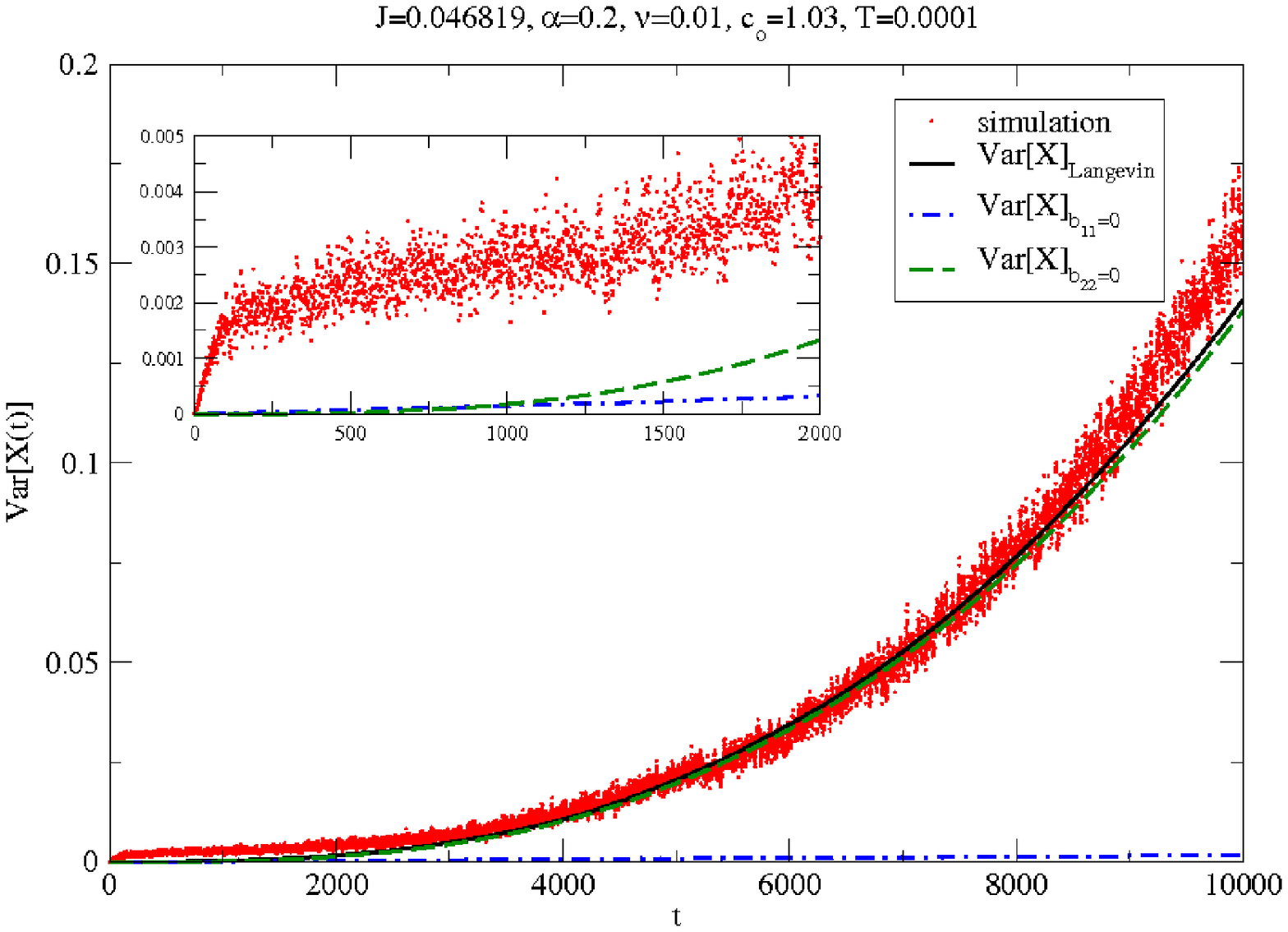}}
\subfigure[]{\includegraphics[width=7cm,angle=270]{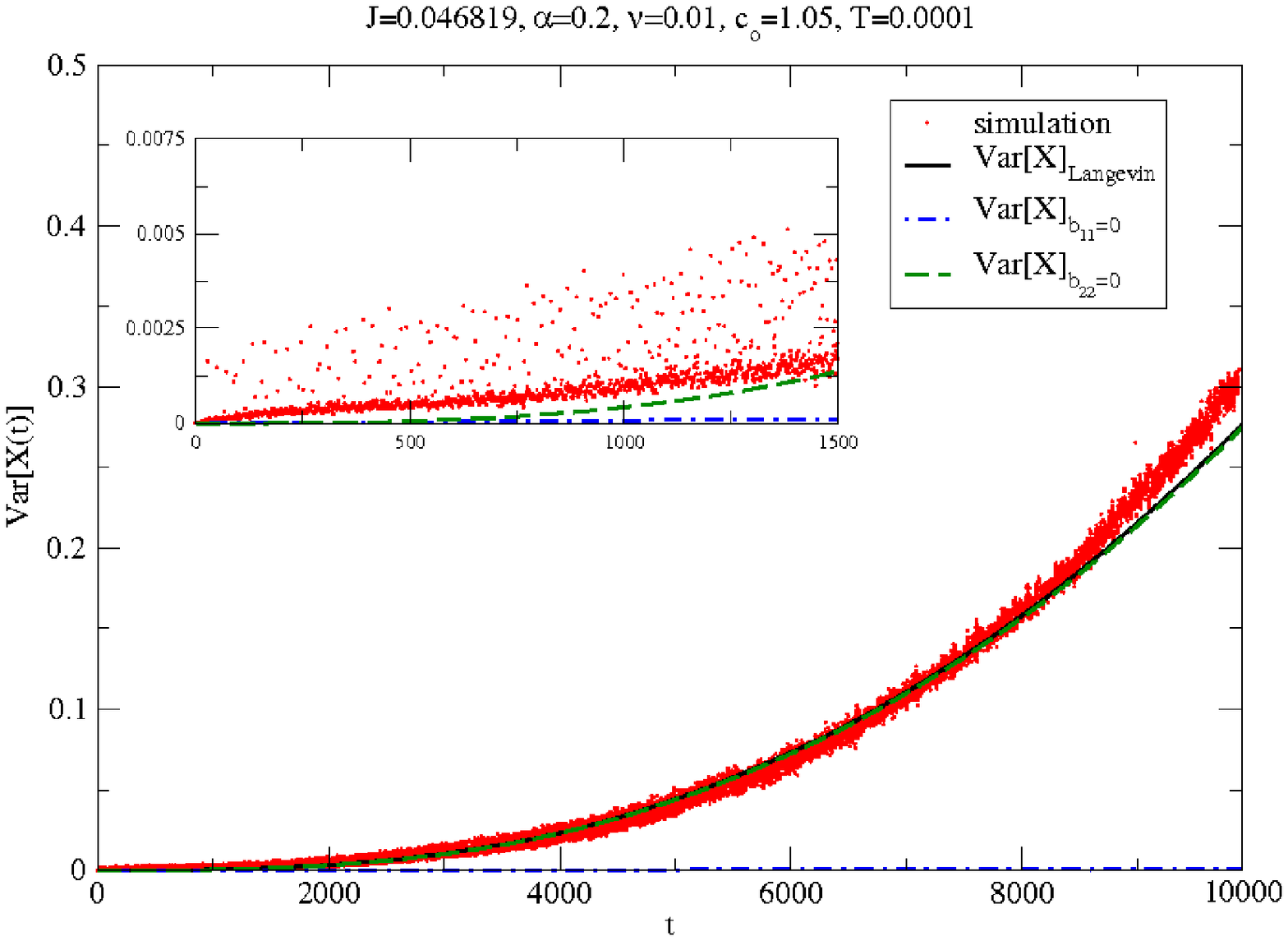}}
\subfigure[]{\includegraphics[width=7cm,angle=270]{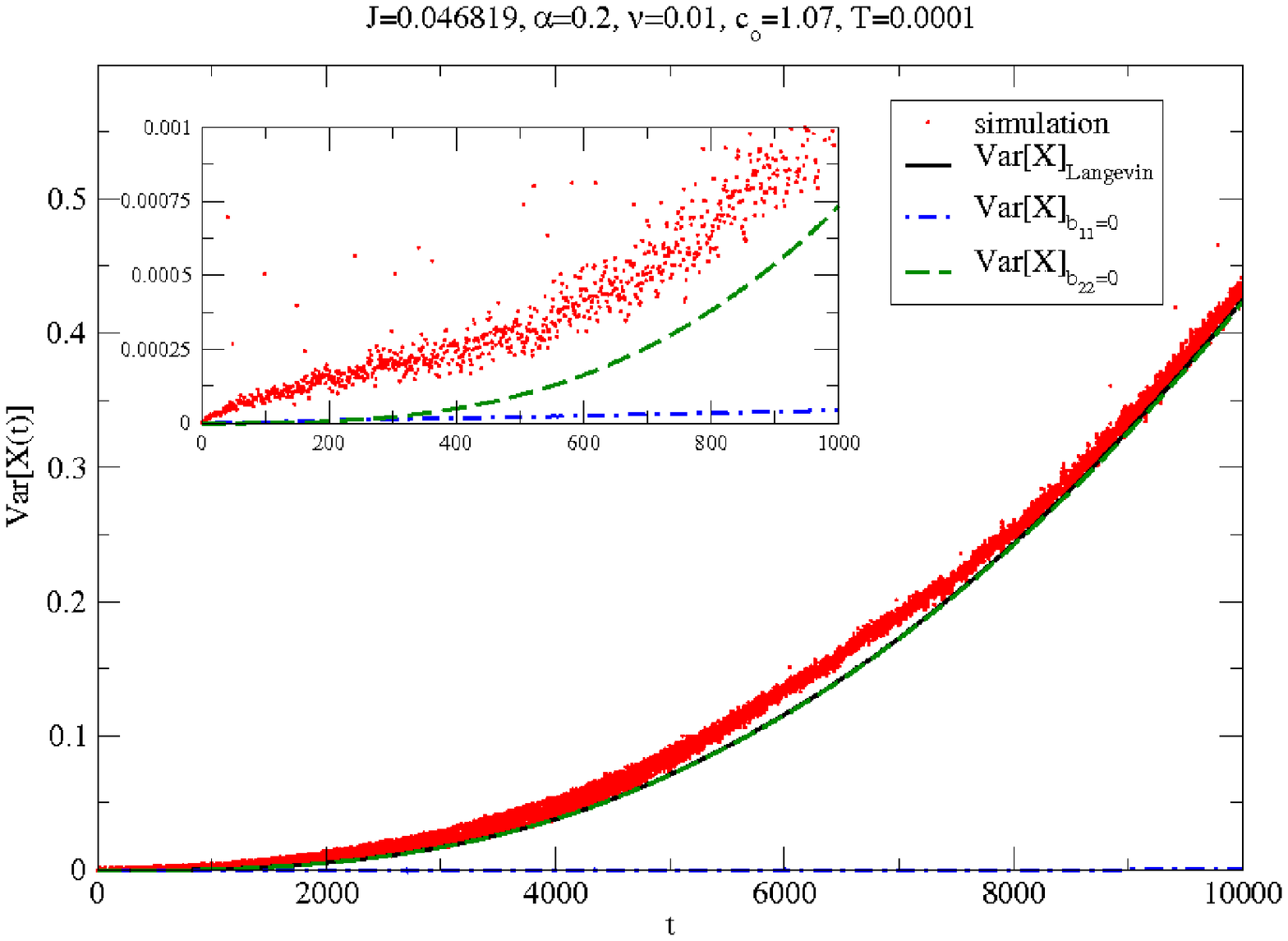}}
\label{soldiffal0.2}
\end{figure}

The time integration of the equations of motion (\ref{discreteeq}) was carried
out by use of the Heun method, which is widely used in the solution of partial differential equations or
difference-differential equations, coupled to either an additive or a
multiplicative noise term \cite{kloeden}. Periodic boundary conditions were used
to enable the running of
long simulation times on systems of typically
$2000$ particles, clearly larger than the cut-off value for the LRI
($100/\alpha$). For a system with LRI, the simulation times increase with 
the square of the system
size $N$. To measure the diffusion of the solitons, over 
$100$ different realizations of the system, which were carried out on a cluster
with 48 processors, were averaged. \\
The results for the position variance $Var[X(t)]$ measured in the simulations and the numerical solution of the Langevin system
(\ref{langevin}) for $\alpha=0.3$ (Fig. \ref{soldiffal0.3}) and $\alpha=0.2$
(Fig. \ref{soldiffal0.2}) \cite{gaididei97} will be compared below.
The purely normal and superdiffusive contributions $Var[X]_{b_{11}=0}$ and $Var[X]_{b_{22}=0}$ are also displayed to demonstrate that for the 
shown velocity range ($c_o=1.03$, $c_o=1.05$, $c_o=1.07$), the soliton diffusion $Var[X]$ is practically equal
to the superdiffusive part $Var[X]_{b_{22}=0}$.
The long-range coupling $J$ and the inverse long-range radius $\alpha$ are
chosen such as to yield a constant value for the sound velocity $c=1.51516$.

The results from both the simulations and Langevin system for $\alpha=0.3$ (Fig.
\ref{soldiffal0.3}) and especially $\alpha=0.2$ (Fig. \ref{soldiffal0.2}),
generally agree
well.
The normal diffusion of the soliton is negligible for solitons with velocities
$c_o\ge1.03$. The smallness of the term $Var[X(t)]_{b_{11}=0}$ shows that the direct action of the noise on the soliton position $X(t)$ is practically negligible.

The superdiffusive mechanism $Var[X(t)]_{b_{22}=0}$ describes the soliton
diffusion very well and the agreement increasingly betters for larger values of the interaction radius $\alpha^{-1}$.
The fluctuations of
$\gamma(t)$ yield a fluctuating drift term $v_d(\gamma(t))$ and thus
a fluctuating $X(t)$ with a stronger than linear  variance $Var[X(t)]_{b_{22}=0}$.
Since the agreement for $\alpha=0.2$ (Fig. \ref{soldiffal0.2}) is even better than for $\alpha=0.3$ (Fig. \ref{soldiffal0.3}), the influence of the
stochastic quantity $\gamma(t)$ on the soliton velocity $v_d$ (where the long-range terms contribute) can be regarded as the driving force of the superdiffusive mechanism.  
\\
The Kac-Baker form keeps the sum over the coupling constants $J_m$ constant
(\ref{Jsum}). Therefore, the almost perfect agreement between
$Var[X]_{b_{22}=0}$, and the simulation
results $Var[X(t)]$ for smaller values of $\alpha$, can be regarded as an effect of the stronger long-range character of the system. 
The normal diffusion $Var[X]_{b_{11}=0}$ for the values of $\alpha$ and $c_o$ in
 figures \ref{soldiffal0.3} and \ref{soldiffal0.2} is only of importance for very small times and yields
a clearly smaller slope than the simulation results. This discrepancy appears in
\cite{arevalo03} for solitons on chains with NNI as well, and it was tackled in \cite{mertens05} by generalizing the
results for the soliton diffusion on the Toda lattice to other anharmonic
chains. It was demonstrated that the  phonon-induced diffusion of the Toda soliton (when it is expressed in 
terms of the soliton characteristics (velocity, amplitude, width) of solitons on FPU-like chains) can be used to explain the discrepancy for the prediction of the normal
diffusion.
\\
In figures \ref{soldiffal0.3} and \ref{soldiffal0.2}, the dominance of the
superdiffusion sets in earlier for larger values of $c_o$. This result can be qualitatively explained by the fact that the soliton dynamics 
depends on the time scale $t_r$ (\ref{gamma}), which is proportional to the damping and the square of the inverse width. 
For large values of $c_o$, the width is small and the reduced time $t_r$ increases faster.
Interestingly enough, the time coordinate $\tau$ 
in the analytical result for the position variance of the KdV solitons \cite{arevalo03} appears with a pre-factor $\lambda$ which is also proportional to the damping and the square of the
inverse width. The dependence of the soliton diffusion on such a time scale
seems very typical and is ultimately a consequence of the universal behavior of the soliton width
in the presence of damping (\ref{gamma}).
\\
To quantitatively see the influence of the long-range radius
$\alpha^{-1}$ and the time scale $t_r$ on the soliton diffusion, an analytical
result for the position variance would have to be deduced.
However, because the dependence of $v_d$ on $\theta$ involves a polygamma
function (\ref{vd}), this is a rather hard task. According to our observation, the
superdiffusive part of the simulation results can always be very well
approximated by a mainly quadratic and cubic time dependence.
This observation agrees with the result of a perturbation theory in \cite{arevalo03}
that the superdiffusion yields a term $\sim t^2$ for the position variance of the KdV soliton.  
\\ 
For solitons with relatively low velocities, the normal diffusion is more
pronounced than for their counterparts on NNI chains for simulation times
$t<10000$. The LRI solitons become very broad for small velocities and appear
more as a collective excitation of the system,
than localized energy pulses (see Fig.\ref{nnwwlrwwvergl} for $c_o=1.01$). The
superdiffusive character for solitons, with for example $c_o=1.01$, appears only for $t>10000$, when $t_r$
reaches larger values. 

\begin{figure}[htbp]
\caption{Soliton diffusion for different soliton velocities ($c_o=1.01$ and
$c_o=1.02$) on a chain with only nearest neighbor interactions ($J=0$, $c=1$) and $\nu_{hy}=0.003$ and $T=5\cdot
10^{-6}$. }
\subfigure[]{\includegraphics[width=7cm,angle=270]{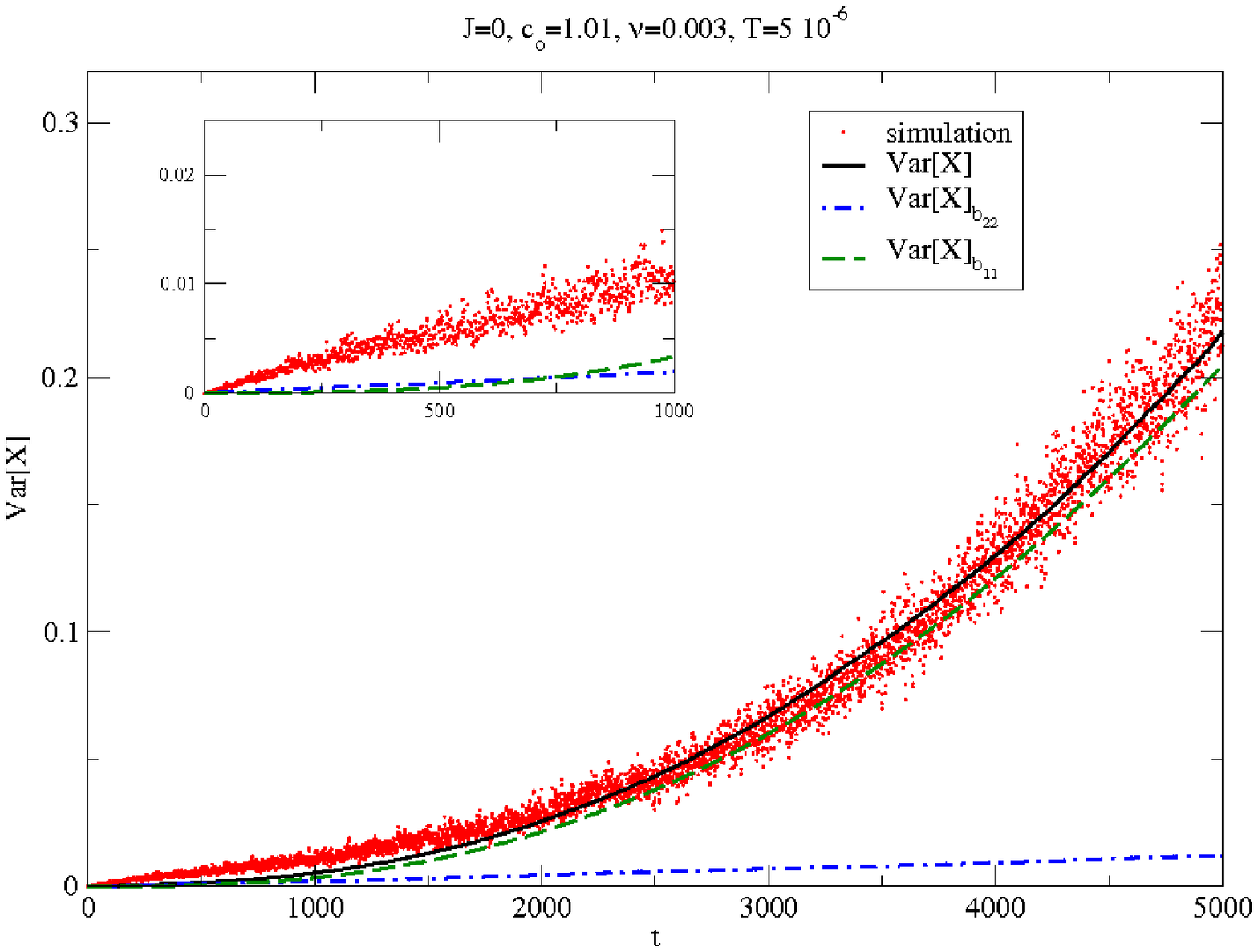}}
\subfigure[]{\includegraphics[width=7cm,angle=270]{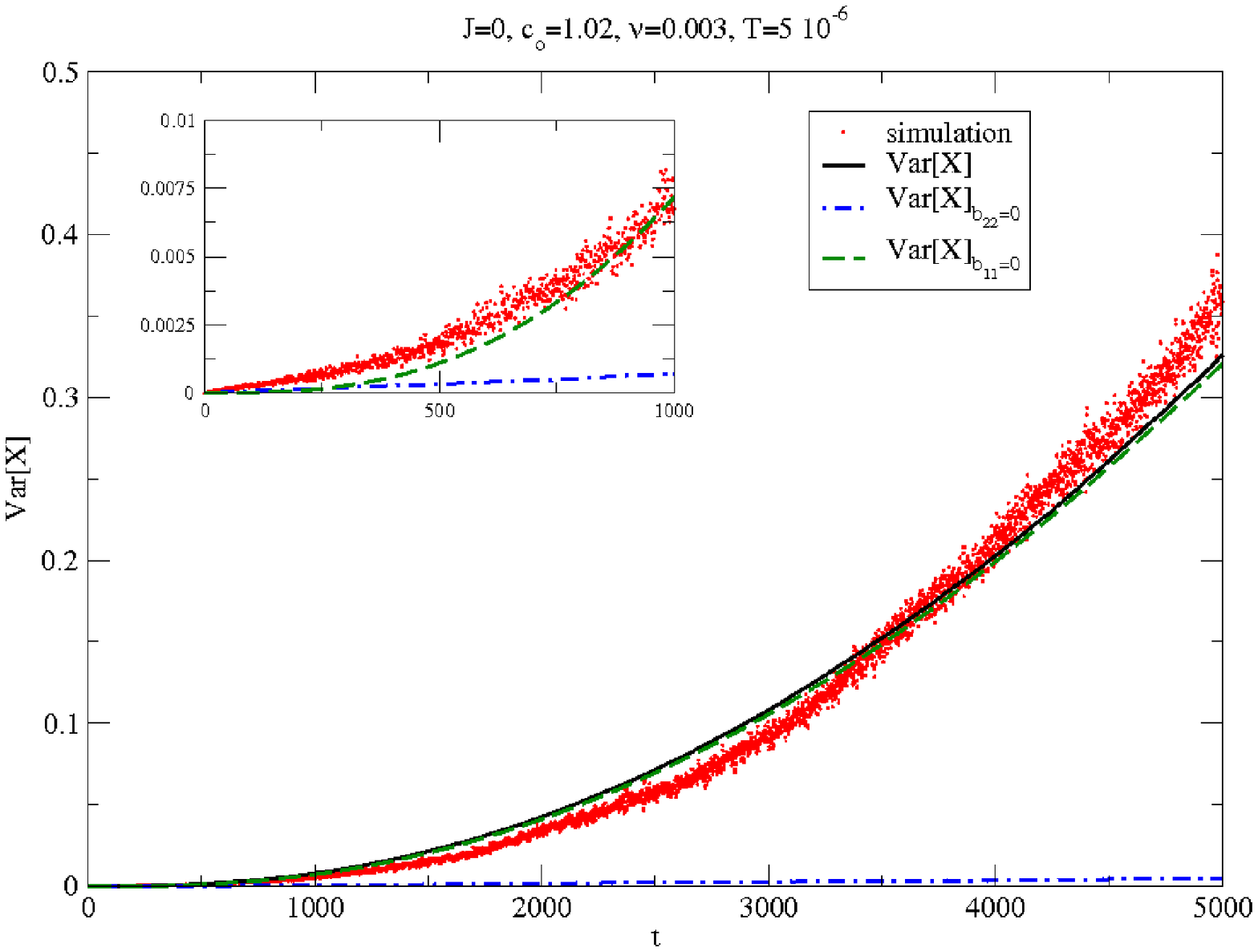}}
\label{soldiffJ0}
\end{figure}

The theory of section IV holds for chains with only nearest neighbor
interactions (by setting $J=0$) as well. In this case, the results agree well to
those of earlier studies \cite{arevalo03,arevalo}. 
The diffusive mechanism is basically the same as for LRI, but because the
discreteness effects are much stronger, the solitons for the nearest neighbor
case are rather narrow and can not manage velocities as high as for LRI.
Therefore, the
normal diffusion is dominant for broad solitons at quite small soliton
velocities, for example $c_o=1.005$. Superdiffusive effects can hardly be seen, because these low-energy solitons
are rapidly destroyed by the noise and damping. For larger values of $c_o$, such
as $c_o=1.01$ and $c_o=1.02$ (in  Figures \ref{soldiffJ0} a and
b), the superdiffusive contribution dominates. 
\\
The numerical results for the soliton diffusion on chains with Kac-Baker
long-range interactions generally suggest that the superdiffusive mechanism is
more pronounced than it is for nearest
neighbor interactions. For LRI, the range of possible velocities where the
superdiffusion dominates is clearly enlarged. The solitons in the LRI case also
have higher energies
and longer life times which because of the quadratic and cubic terms in the
time-dependence of $Var[X(t)]$, lead to the dominance of the superdiffusion.

\section{Small-noise expansion and long-time limit}
In order to gain some deeper insight in the diffusion mechanism of the
long-range solitons and the time-dependece of $Var[X(t)]$, we develop a
small-noise expansion of the Langevin set 
(\ref{langevin}). The most important step in the analytical treatment is the
rather complicated dependence of the velocity $v_d$ on the inverse width
$\gamma(t)$. An expansion of the polygamma function $\Psi^{\prime
\prime}(\theta)\approx -\frac{1}{\theta^2}-\frac{1}{\theta^3}-\frac{1}{2
\theta^4} +\frac{1}{6 \theta^6}$ for large $\theta=\kappa/2 \gamma$ \cite{abramowitz} is
already rather appropriate for typical values of
$\theta(t=0)\in [1.5,3]$ for the considered long-range solitons with
$c=1.51516$ in Fig. \ref{soldiffal0.3} and Fig. \ref{soldiffal0.2}. This approximation is
even for smaller values of $\theta$ 
justified after some time $t > 1/\gamma^2 \nu_{hy}$, because of the
damping-induced growing of the soliton width $\gamma^{-1}$.
It can be shown that the soliton velocity $v_d$ depends in this approximation
only on an additional 
term which is proportional to $\gamma^2$:
\eas
v_d&=&\sqrt{1+\frac{8}{9} A_o \gamma(t)-2(c^2-1) \bigg[\theta+\theta^2+\theta^4
\Psi^{\prime \prime}(\theta) \bigg]} \nonumber \\
&\approx& \sqrt{1 +\frac{8}{9} A_o \gamma(t)-2 (c^2-1) \big[
-\frac{1}{2}+\frac{1}{6 \theta^2}\big]} \nonumber \\
&\approx& c \sqrt{1+\frac{8}{9} \frac{A_o}{c^2} \gamma(t)-\frac{4}{3}
\frac{(c^2-1)}{c^2 \kappa^2} \gamma^2} \nonumber \\
&\approx& c+\frac{4}{9} \frac{A_o}{c} \gamma(t)-\big(\frac{8}{81}
\frac{A_o^2}{c^3}+\frac{2}{3} \frac{(c^2-1)}{c \kappa^2} \big) \gamma(t)^2 ~.
\eae 

\begin{figure}[htbp]
\caption{Simulation results for the position variance of a soliton
with 
$c_o=1.03$ on a chain with $J=0.046819$, $\alpha=0.2$, $c=1.51516$,
$\nu_{hy}=0.01$ and $T=0.0001$. The results agree well with the numerical
solution of the Langevin system (\ref{langevin}). The
result from the small-noise expansion reproduces the numerical solution of the
Langevin system qualitatively well.}
\subfigure[]{\includegraphics[width=7cm,angle=270]{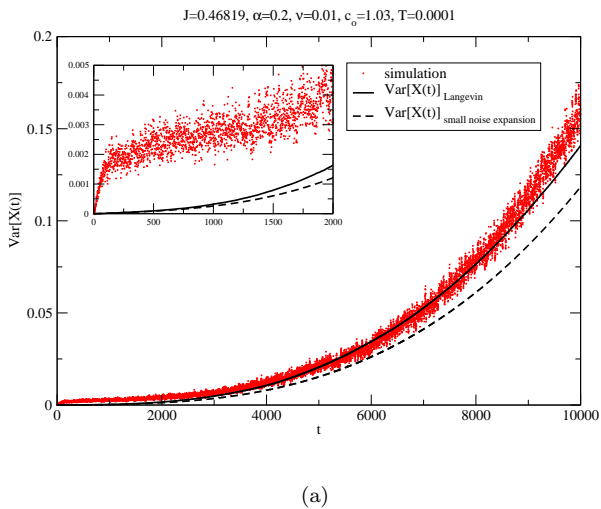}}
\label{snexp}
\end{figure}

The small-noise expansion of the Langevin set (\ref{langevin}) follows reference \cite{gardinier} and was already successfully applied to the problem of the diffusion of
low-velocity solitons in the FPU system without long-range interactions \cite{arevalo03}. We seek an asymptotic solution of the form
\eas
\gamma(t)=\gamma^{(0)}(t)+\epsilon \gamma^{(1)}(t)+ ...\\
X(t)=X^{(0)}(t)+ \epsilon X^{(1)}(t)+... 
\eae
where $\epsilon$ is a small parameter which is formally introduced to consider the influence of the noise terms as small perturbations ($\sqrt{D^{hy}}\sim \epsilon$).
The contribution of the drift term in the different orders of $\epsilon$ are calculated following the rule
\eas
a_i(\gamma(t))&=&a_i\bigg(\gamma^{(0)}+\sum_{m=1}^{\infty} \epsilon^m
\gamma^{(m)}(t) \bigg)\nonumber \\&=&a_i^{(0)}(t)+\epsilon \gamma^{(1)}(t)\frac{d
a_i(\gamma^{(0)}(t))}{d
\gamma^{(0)}(t)}+
... ~~.
\eae
With $a_2=v_d$, the equations in the order $\epsilon^0$ read:

\eas
d \gamma^{(0)}(t)&=&-0.8 \nu_{hy} \gamma^{(0)}(t)^3 dt
\\
d X^{(0)}(t)&=&\bigg(c+\frac{4}{9} \frac{A_o}{c} \gamma(t)\nonumber \\&-&\bigg(\frac{8}{81}
\frac{A_o^2}{c^3}+\frac{2}{3} \frac{(c^2-1)}{c \kappa^2} \bigg) \gamma(t)^2  \bigg) dt
\eae
and describe the damped soliton. The first corrections due to the noise terms appear
in the next order $\epsilon^1$:
\eas
d \gamma^{(1)}(t)&=&-2.4 \nu_{hy} \gamma^{(1)}(t)\gamma^{(0)}(t)^2
dt\nonumber \\&+&\sqrt{\frac{3}{5}}\frac{\sqrt{D} \gamma^{(0)}(t)^{\frac{3}{2}}}{A_o v} dW_1\nonumber \\
\\
dX^{(1)}(t)&=&\gamma^{(1)}(t) \bigg(
c+\frac{4}{9} \frac{A_o}{c} \nonumber \\&-&\bigg(\frac{16}{81}
\frac{A_o^2}{c^3}+\frac{4}{3} \frac{(c^2-1)}{c \kappa^2} \bigg) \gamma^{(0)}(t)
\bigg)dt\nonumber \\&+&\sqrt{\frac{1}{6}+\frac{\pi^2}{180}}\frac{\sqrt{D}}{ \sqrt{\gamma^{(0)}} A_o v}
dW_2~.
\label{x1}
\eae
If we substitute the result for $\gamma^{(1)}$ and
$\gamma^{(0)}=\gamma_o/\sqrt{1.6 \nu_{hy}
\gamma_o^2 t+1}$ into (\ref{x1}), we can calculate the 
first-order expression for the variance of the soliton position due to the noise
\eas
Var[X_1(t)]=\lim_{s \rightarrow t} <X_1(t) X_1(s)>~.
\eae
This is the first order result for the soliton position variance $Var[X(t)]$ when we
finally set $\epsilon=1$ which is justified if the noise in the system is
sufficiently small. Since the small-noise expansion yields a very lengthy result for the position
variance, which depends on the time scale $t_r=1.6 
\nu_{hy} \gamma_o^2 t$, we prefer to demonstrate the good agreement just for one
typical example (Fig. \ref{snexp}). 

Further comparisons reveal that the analytical result for $Var[X(t)]$ begins to
fail for high velocities $v$ close to the critical velocity $v_c$. The reason is the
restriction to the first order corrections. For high-velocity solitons, the
width is quite narrow and the fluctuations of the width are strong
and can no longer be well described by $\gamma^{(1)}(t)$. 
\\
The long-time limit for the soliton diffusion result in the small-noise
expansion predicts an asymtotic soliton diffusion where the underlying soliton
shape determines only the numerical constant in
\eas
Var[X(t)]_{inf}=\frac{1}{64} \frac{D^{hy}}{c^2 v^2 \nu_{hy}^{1.5}} t^{3/2}~.
\eae
We can not prove this prediction in our simulations because the result Var[X(t)]
for the small-noise expansion reaches
the asymptotic $t^{3/2}$ time-dependence after very long times. It would take times larger than $10^{6}$ for the soliton of
Fig. \ref{snexp}. For these times, the
soliton is not a localized energy pulse but smeared out over the system and the
soliton velocity $v$ approaches the velocity of sound $c$. Therefore,
we do not assume the long-time limit to play a role in energy transfer
mechanisms. As long as the soliton is not too broad, the time
dependence of the soliton diffusion can be well approximated by a quadratic
and cubic term in $t$.

\section{Conclusions}
It was observed that the diffusion of solitons on chains with Kac-Baker long-range interactions is in many respects very similar to their counterparts on chains with nearest
neighbour interactions. A theory which describes both cases very well, was
derived here. Especially the dependence of $\gamma(t)$ on $t_r$, and the superdiffusive mechanism
with the mainly quadratic contribution in time to the position variance,
in both cases, gives rise to the conjecture that superdiffusion is a generic
feature for non-topological solitons.
\\
In the case of long-range interactions, the solitons can adopt rather high
velocities and it was observed that 
the superdiffusive mechanism of the Langevin system, which was derived in the framework of the applied CV theory for the non-local Boussinesq equation with noise
and damping terms, describe them well.  
The prediction of the derived Langevin set for the superdiffusion of the soliton is especially accurate for small values of $\alpha$, where solitons (for a certain velocity
$c_o$) possess more energy than for larger values of $\alpha$. It is not surprising that the superdiffusion dominates for massive solitons because the direct influence of the
noise on the soliton position is less important. 
The superdiffusion is more present for solitons with LRI, because the possible velocity range and the lifetime of the solitons are distinctly enlarged.
For a chain with nearest neighbor coupling, solitons could only be described to
rather moderate velocities, where the normal diffusion is more present. 
\\
In the framework of a small-noise expansion for the Langevin system
(\ref{langevin}), we derived
an analytical result for the position variance $Var[X(t)]$ which agreed well
with the simulations when the soliton velocities are not too large. This result 
depends in a quite complicated way on the time scale $t_r$. But the result confirms the 
former observation that the superdiffusion yields contributions in $t^2$ which 
turn into a $t^{3/2}$ dependence for long times when the soliton is rather
broad with a velocity $v$ close to the sound velocity $c$.    
\\
For small times, the soliton with LRI showed a normal diffusion which is larger than the result $Var[X(t)]_{b_{11}=0}$ of the CV theory.
The problem in determining the normal diffusion of the soliton may be two-fold.
In the case of a chain with nearest neighbor coupling
\cite{arevalo03,mertens05}, it was established that the phonons have a significant 
contribution to the diffusion of solitons, especially for low-velocity solitons.
This result could possibly be similar for LRI solitons. When, as in this paper,
higher velocities are examined, the KdV equation is no longer suitable
describing the situation well.
Instead, the Bq equation, where a second-order time derivative of the wave
function appears, which results in ODEs for the CVs where the soliton
position appears quadratic ($\dot{X}^2$), has to be used. To approximate this
result by a Langevin-type equation, the square root of the r.h.s. of (\ref{rewritten})
had to be expanded because this procedure yields the correct velocity $v_d$
without noise. However, with this procedure and by substituting $v_d$ with a constant value $v$, the result has the
tendency to underestimate
the influence of the noise on the CV $X$. Thus, it is difficult to decide wether
it is the phonons or the applied approximations, which are responsible for the observed discrepancies. 
\\
Nevertheless, the predicted superdiffusive behavior of localized energy pulses
in systems with long-range forces can be observed for long enough times.  
Moreover, the long-range interactions lead to much larger soliton energies and
lifetimes, so that the soliton diffusion could be an observable effect in real systems.


\end{document}